 \documentclass[final,5p,times,twocolumn,authoryear]{elsarticle}

\usepackage{url}

\usepackage{amssymb}
\usepackage{lipsum}

\journal{Nuclear Physics B}

\begin{document}

\begin{frontmatter}



\title{What Makes A Discovery?}


\author[first,second,third]{Dan Hooper}
\affiliation[first]{Fermilab, Theoretical Astrophysics Department}
\affiliation[second]{University of Chicago, Kavli Institute for Cosmological Physics}
\affiliation[third]{University of Chicago, Department of Astronomy and Astrophysics}
\date{\today}


\begin{abstract}

In this contribution to the proceedings of the 182nd Nobel Symposium, I reflect on the concept of ``discovery'' as it is used by physicists and astronomers. In particular, I comment on how the scientific community distinguishes discoveries from propositions that are supported only by lesser forms of evidence, emphasizing the social nature of this process and remarking on the subjective factors that go into making such judgements. I advocate for an approach that is intentionally Bayesian in nature, in which individuals are encouraged to evaluate and publicly state their priors and to update them systematically. I close by applying these practices to the case example of the Galactic Center Gamma-Ray Excess.

\end{abstract}






\end{frontmatter}




\section{Introduction}

I recently had the pleasure of attending a Nobel Symposium on the topic of dark matter. At that meeting, I gave a talk describing the status of the Galactic Center Gamma-Ray Excess. Given that you can find my thoughts on that topic elsewhere~\citep{Hooper:2022bec,Xu:2023zyz,Keith:2022xbd}, I decided to write these proceedings about my reflections on another series of discussions that took place at that meeting.

Over the course of the Nobel Symposium, several of the participants had a series of lengthy conversations in which we discussed the following question: What would have to occur before we, as a scientific community, would be willing to declare that we had discovered the nature of dark matter.\footnote{By ``the discovery of the nature of dark matter'' I mean the discovery of the particles that make up the dark matter (or the majority of the dark matter). I consider the existence of dark matter to have been firmly established for many years. For a historical account, see~\cite{Bertone:2016nfn}.} Most of these discussions were practical in nature, with various participants suggesting examples of hypothetical experimental signals that might one day rise to the level of an authentic discovery. There was little, however, that was particularly systematic about these discussions. Different people in the conversation -- all of whom were experts in their respective fields -- seemed to apply different kinds of criteria to the question at hand, and otherwise approach the question in a diversity of ways. I left the meeting feeling intrigued by the very concept of discovery, including the question of how we as a community decide what has and has not been discovered and, more broadly, in the role that this concept has played in the history of science, and especially in that of physics and astronomy.

\section{What Makes a Discovery a Discovery?}

The scientific method is often described in almost cartoonishly simplistic terms, to such a degree that it barely resembles the activities of science as they are practiced in the real world. In this common caricature, the scientific method follows a sequence of well-defined steps, such as posing questions, constructing hypotheses, making predictions, testing predictions, and rejecting or refining hypotheses. 
As it is actually practiced, however, the scientific method does not resemble the steps of a recipe. Instead, there is a great deal of complex social behavior that goes into determining how the scientific community responds to new information. This complexity is not necessarily a bad thing; in some cases, there can be considerable benefits to the so-called ``wisdom of the crowd.'' If we are not thoughtful about how these collective judgements are formed, however, we run the risk of sub-optimally allocating our time and other resources. After all, our collective judgements impact which experiments are to be funded and ultimately carried out, as well as which ideas are to be explored and to what depth. If made poorly, these judgements can delay scientific progress, and otherwise cause us to learn less about our universe than we potentially could have.

Perhaps the most well-known work on the process of science as it is actually practiced is {\it The Structure of Scientific Revolutions} by the historian and philosopher Thomas Kuhn~\citep{Kuhn1962}. In his book, Kuhn introduced the concept of a ``paradigm shift,'' and argued that science does not progress through the steady accumulation of knowledge, but instead undergoes alternating periods of ``normal'' and ``revolutionary'' activity. During periods of normal science, data is collected and new facts are learned while operating within the boundaries of an established paradigm. During scientific revolutions, in contrast, old ideas are discarded as new ones are built up and adopted, resulting in the creation of new paradigms. There is no objective criteria or formula for determining when a new paradigm will be established. Furthermore, such revolutions do not occur at the individual level, but instead take place among communities of scientists, working across a given field or subfield of research. Paradigm shifts are thus fundamentally social in nature.

When I was a graduate student, I found myself under the impression that there existed a clear line of demarcation between those scientific facts that had been discovered and those provisional propositions that were supported only by lesser forms of evidence. I recall us all agreeing at that time that the top quark had been discovered several years earlier~\citep{D0:1995jca,CDF:1995wbb}, but that the early evidence of the Higgs boson from LEP II~\citep{ALEPH:2000zfn,L3:2000qpr} did not rise to that standard.\footnote{Both of these judgements would hold up well to future data.} It was striking just how clear-cut and unambiguous this distinction appeared to be.

In the field of particle physics, the word discovery  is sometimes treated as a synonym for one or more measurements that are discrepant with the predictions of the null hypothesis at a level of at least 5 standard deviations  (``evidence'' is similarly reserved for measurements that are discrepant by at least 3 standard deviations). Much like the step-by-step caricature of the scientific method, this simple definition of what it means to make a discovery does not hold up to historical scrutiny. In practice, the scientific community takes much more than measures of statistical significance into account when deciding how to respond to new data. By scrutinizing and systematically refining the processes by which we collectively make such judgements, we could potentially improve our ability to effectively allocate resources, and otherwise increase the rate at which we are able to make scientific progress.

\section{On Statistics and Systematics}

As a branch of mathematics, statistics offers a means of obtaining objective and well-defined answers to certain precisely posed and well-defined questions. When confusion or ambiguities arise regarding a question of statistics in the physical sciences, it is usually not due to mathematical error. Rather, it is more often the case that someone has asked one question but answered another.

There exist many examples of experimental or observational anomalies that could be said to have met the traditional $5\sigma$ discovery threshold, and yet that are not generally regarded as discoveries by the scientific community. The annual modulation of the event rate reported by the DAMA Collaboration, for example, has a quoted significance of $8.9\sigma$~\citep{DAMA:2010gpn}, and yet very few of us think that this experiment is detecting particles of dark matter. Anomalies and tensions regarding the rate of Hubble expansion~\citep{DiValentino:2021izs}, and the muon's anomalous magnetic moment~\citep{Venanzoni:2023mbe,Athron:2021iuf} are also at or near the $5\sigma$ threshold, and yet have not attained the status of discoveries.

There are many reasons for why a given anomaly might not be treated as a discovery of new physics, even in the presence of very high statistical significance. In some generality, an excess or other anomaly could appear in a given data set for any combination of the following reasons:
\begin{enumerate}
    \item{The result could be a statistical fluctuation.}
    \item{The systematic uncertainties could be underestimated.}
    \item{The measurement could be in error.}
    \item{There could be an underestimated trials factor.}
    \item{The anomaly could be a real signal that is the result of new physics.}    
\end{enumerate}

In the case of the signal reported by DAMA, the statistical significance is so high as to essentially rule out options 1 and 4, while option 5 is all but excluded by the null results of other dark matter experiments. This suggests that this anomaly is likely to be a consequence of some unknown systematic issue, or some kind of measurement error. In the case of the Hubble tension, all five of the possibilities listed above could plausibly be at play. The same is true for the muon's anomalous magnetic moment, although systematic uncertainties related to the the Standard Model prediction of this quantity seem perhaps most likely to be behind this tension (for a review, see~\citealp{Aoyama:2020ynm}).

As illustrated by these and many other examples, it is often unclear whether experimental backgrounds or other systematic uncertainties associated with a measurement are understood well enough to support a claim of discovery. Such issues can be mitigated in cases where the measurements under consideration provide a diverse collection of complementary information. Consider, for example, a measurement of a single number -- perhaps a rate or a branching fraction -- that is $5\sigma$ discrepant with the value predicted by the Standard Model. It would be relatively easy in such a case to conceive of reasons for the discrepancy that do not involve new physics. Contrast this with a measurement that includes multiple kinds of events, or a distribution of those events, or an image, or other such complex information. Although such a signal could, in principle, arise from a statistical fluctuations or some kind of unknown background, this would require something of a conspiracy of factors. Such signals are especially compelling when the features of the measured data are easily explained by one or more well-motivated models of new physics.  

In the following two sections, I will focus on two specific factors that can influence our collective assessment of experimental anomalies: the counting of trials factors, and theoretical priors.

\section{Trials and Trials of Trials}

In 2016, an excess of diphoton events with an invariant mass of approximately 750 GeV was reported at the Large Hadron Collider (LHC). Initially, the (local) significance of this 750 GeV excess was reported to be 3.9$\sigma$ at ATLAS~\citep{ATLAS:2016gzy} and 3.4$\sigma$ at CMS~\citep{CMS:2016xbb}, corresponding to chance probabilities of 0.00005 and 0.0003, respectively. Even after accounting for the ``look elsewhere'' effect arising from the fact that these analyses searched for analogous signals across a number of different invariant mass values (constituting different ``trials'')~\citep{Gross:2010qma}, the odds of obtaining such a signal at one or more mass was found to be 0.017864 and 0.0548, corresponding to global significances of 2.1$\sigma$ and $1.6\sigma$, respectively. Treating the ATLAS observation as a confirmation of the CMS excess, the probability of obtaining such a result (in the absence of new physics) is only $0.0003 \times 0.017864=5\times 10^{-6}$, corresponding to approximately 4.6$\sigma$ significance, and only slightly below the traditional 5$\sigma$ threshold for discovery. To the disappointment of many, this excess began to decline as ATLAS and CMS collected more data, revealing it to have been nothing more than a statistical fluctuation~\citep{ATLAS-CONF-2016-059,CMS-PAS-EXO-16-027}.

Although this excess generated a great deal of excitement at the time,\footnote{Roughly 500 papers on this excess appeared on the arXiv in the months following its announcement.} I don't know of anyone who thought that it was especially likely that this excess was due to new physics. I certainly never heard anyone express the view that the odds of this were anywhere close to $1-(5\times 10^{-6}) \approx 99.9995\%$. Even the most enthusiastic proponents of this excess treated this likelihood to be on par with that of a coin toss. Given its quoted statistical significance, I find it interesting to ask why this was the case. 


I know of no reason to suspect that there were any measurement errors of consequence involved in the appearance of the 750 GeV excess. Instead, I would argue that this anomaly was more likely the result of a statistical fluctuation combined with a large and somewhat difficult to quantify trials factor. After all, many thousands of different searches for new physics have been carried out at the LHC, and thus anomalies with a statistical probability of $\mathcal{O}(10^{-3})$, or even $\mathcal{O}(10^{-4})$, should appear in this data set from time to time. If we take this factor to be $10^4$, for example, it would increase our estimated odds of observing an anomaly as unlikely as the 750 GeV excess to $\sim 5 \times 10^{-6} \times 10^4 \sim 5\%$. While such a result could be seen as a bit unlikely, it is far from implausible.

\section{Theory Bias or Theory Guidance?}

As a very different example, consider the scientific community's reaction to the LHC's discovery in 2012 of ``a new particle with a mass of 125 GeV'' (aka, the Higgs boson). When this result was first announced, the CMS and ATLAS collaborations reported 4.9$\sigma$ and 5.0$\sigma$ evidence in favor of the existence of this particle, respectively. Almost immediately, most of us in the particle physics community began to refer to this as the ``discovery of the Higgs boson.''

The response to this announcement was strikingly different from that of the 750 GeV diphoton excess. This was in part due to the higher statistical significance of the Higgs discovery, overcoming the look elsewhere penalty associated with even a large trials factor. I would argue, however, that there was another important factor at play in this situation. Namely, the scientific community fully expected that the LHC would discover the Higgs boson, and that it would have properties (production rates, branching fractions, etc.) that were broadly consistent with those reported by ATLAS and CMS in their 2012 announcement. In contrast, no one was expecting that the LHC would discover a new 750 GeV particle with the couplings and other characteristics that would be required to generate the reported diphoton anomaly. In other words, most of us evaluated the {\it prior probability} of the Higgs boson to be much greater than that of something capable of generating the 750 GeV excess. 

Some of my colleagues will surely object to one taking these kinds of expectations into account when evaluating the likelihood that a given anomaly or excess is the consequence of new physics. In the following section, I will take this opportunity to explain why I respectfully disagree. 



\section{Intentional Bayesianism}

In contrast to mathematics, the pursuit of science cannot rely entirely on the logical techniques of deduction and induction. Instead, the scientific method can be thought of as an application of abduction, defined as inference to the best explanation~\citep{sep-abduction}. A well known example of abductive reasoning is the process of Bayesian inference, as formulated in terms of Bayes' theorem.

Bayes' theorem can be used to determine the probability that a given theory is true, taking into account both prior knowledge and newly acquired data. Given a set of data, Bayes' theorem states that the (posterior) probability of a given theory can be expressed as follows:
%
%
\begin{equation}
\label{bayes}
P(T_i|D) = \frac{P(D|T_i) \, P(T_i)}{P(D)}. 
\end{equation}
The quantity $P(D|T_i)$ is the likelihood that the data set $D$ would be obtained if the theory $T_i$ is true, $P(T_i)$ is the prior probability that the theory $T_i$ is true, and $P(D) = \sum_i P(D|T_i) \, P(T_i)$ is a normalization factor.

Bayes' theorem is a mathematical theorem, and is unambiguously true. That being said, the results of the equation above ({\it i.e.}~the posterior probabilities) depend critically on one's priors. Unlike the other quantities appearing in Bayes' theorem, the priors are fundamentally subjective in nature; there is no algorithm or other objective criteria that can be used to establish these priors. Instead, the values of $P(T_i)$ are a matter of judgement. In practice, one's priors can take into account a variety of factors, including an assessment of a theory's naturalness, fine-tuning, motivation, and overall plausibility.

Given the fundamentally subjective nature of Bayesian priors, different individuals will assign them different values. One might think that this would make Bayes' theorem useless, but this turns out to not be the case. Importantly, as more data is accumulated, the posterior probabilities of different individuals will converge, even if they start out with very different priors. Bayesian inference thus makes it possible to build consensus, at least in the presence of sufficient data. 

In our ordinary behavior, we are all Bayesians -- at least informally. We are all more skeptical of information that surprises us, and are more likely to accept information that confirms our expectations. This is Bayesian inference at work. Whereas a critic might point to this as the nefarious influence of theoretical bias, I call it the appropriate and rational application of abductive reasoning.


In practice, there are many instances in which human beings fail to properly apply Bayesian principles. In particular, I have noticed a number of instances in which an individual's stated position on a given proposition (i.e., their posterior probability) does not change as new and relevant information is obtained. What I speculate is going on in these instances is that the person in question has in mind -- or in their gut -- a vague notion of the likelihood that the proposition under consideration is true, and that they then proceed to subconsciously reverse engineer their priors such that they obtain something close to their intuitively desired posterior probability. In other words, they are changing their priors in order to counteract the impact of the new information. This failure mode of human psychology can be counteracted -- or at least mitigated -- by one reflecting upon and clearly stating their initial priors, effectively preventing them from retroactively editing them as new information comes to light. To my surprise and sometimes frustration, I find that many of my colleagues are reluctant to quantitatively state their estimated priors (or their posteriors, for that matter). In my opinion, we would all reach more reliable conclusions -- both individually and as a scientific community -- if we were more thoughtful and deliberate in assessing our own Bayesian priors.

\section{The Application of the Galactic Center Gamma-Ray Excess}

In this section, I will use the signal known as the Galactic Center Gamma-Ray Excess to illustrate some of the points that I have been trying to make in these proceedings. This excess, as found in the data collected by the Fermi Telescope, is highly statistically significant and exists for all existing models of the astrophysical diffuse emission and known point sources~\citep{Cholis:2021rpp,DiMauro:2021raz} (for early work, see~\citealp{Goodenough:2009gk,Hooper:2010mq,Hooper:2011ti,Abazajian:2012pn,Hooper:2013rwa,Gordon:2013vta,Daylan:2014rsa,Calore:2014xka,Fermi-LAT:2015sau}). The intensity of this excess emission peaks near the direction of the Galactic Center, and is clearly spatially extended both above and below the Galactic Plane. The Galactic Center Gamma-Ray Excess has been interpreted as a possible signal of annihilating dark matter, but could also be generated by a large population of faint millisecond pulsars. 


Before considering any gamma-ray or other related data, I will begin by stating some of my priors, as I recall them to have been before the original observation of the gamma-ray excess (circa 2008): 
\begin{enumerate}
\item{There is a probability of 0.4 that the main component of the dark matter is a thermal relic.}
\item{Contingent on point 1 being true, there is a probability of 0.2 that the mass of the dark matter particle is in the range of 10-100 GeV.}
\item{Contingent on points 1 and 2 being true, there is a probability of 0.7 that the annihilation cross section of the dark matter particle is within a factor of 5 of $\sigma v = 2 \times 10^{-26} \, {\rm cm}^3/{\rm s}$ (the value predicted for a velocity-independent thermal relic that makes up all of the dark matter).}
\end{enumerate}

I based these priors on a variety of considerations, and another reasonable and well-informed individual might have come up with different values at that time. For example, my prior for point 1 is based on a subjective assessment of minimality and plausibility. If someone thought that this probability should instead be closer to 0.05 or to 0.7, I would not have objected very much to that view. I doubt that many individuals with a background in particle cosmology would have estimated this probability to be very far outside of this range, however. Similarly, in order for a thermal relic to freeze out with the measured density of dark matter without ruining the successful predictions of Big Bang nucleosynthesis, it must have a mass in the range of $\sim 10^{-3} - 10^5 \, {\rm GeV}$~\citep{Boehm:2013jpa,Nollett:2014lwa,Griest:1989wd}. Given that these and other constraints tend to be easier to satisfy near the middle of this range, I came up with the estimated probability of $\sim 0.2$ for the mass range of 10-100 GeV. Taken together, these numbers yield an estimated prior probability of $\sim 0.4 \times 0.2 \times 0.7 =5.6\%$ that annihilating dark matter would produce a signal from the Galactic Center that is qualitatively similar to that observed by the Fermi Telescope. 

Next, I will state my estimated priors as they relate to the astrophysical backgrounds (from pulsars) that could potentially produce a signal similar to that of the Galactic Center Excess:
\begin{enumerate}
\item{There is a probability of 0.2 that there are enough pulsars in the Inner Galaxy to produce a gamma-ray signal with an intensity comparable to that of the observed excess.}
\item{Contingent on point 1 being true, there is a probability of 0.5 that the pulsars' spatial distribution would be qualitatively similar to that of the observed gamma-ray excess (as opposed to being dominated by the central stellar cluster, or significantly elongated along the Galactic Plane, for example).}
\end{enumerate}

Taking all five of these priors together, I would have estimated at the time that there was a 5.6\% probability that dark matter would generate a signal similar to that of the Galactic Center Gamma-Ray Excess, a 10\% probability that pulsars would generate a similar signal, and an 85\% chance that no such signal would appear in the Fermi data. When this signal was identified, and as its spectrum and angular distribution were measured, I came to evaluate $P(D|T_{\rm DM}) \sim P(D|T_{\rm pulsars}) \sim 1$ and $P(D|T_{\rm Neither}) \sim 0$. In other words, this signal could be explained by either dark matter or pulsars, but required at least one or the other to be present. This data, therefore, led me to arrive at posterior probabilities that were approximately given by $P(T_{\rm DM}|D) \sim 0.36$, $P(T_{\rm pulsars}|D) \sim 0.64$, and $P(T_{\rm neither}|D) \sim 0$.

In the years that followed, new data became available which was relevant to interpretations of this signal. Some of my additional priors (circa 2013) were:
\begin{enumerate}
\item{If the Galactic Center Gamma-Ray Excess is generated by dark matter, there is a probability of 0.7 that a direct detection signal would be observed by 2023 (by experiments such as LZ and XENONnT).}
\item{If the Galactic Center Gamma-Ray Excess is generated pulsars, there is a probability of 0.7 that at least several of the brightest gamma-ray pulsars in the Inner Galaxy would be detected by 2023 (by Fermi).\footnote{This not being the case would require the members of the Inner Galaxy pulsar population to be systematically less luminous than pulsars observed elsewhere~\citep{Hooper:2015jlu,Hooper:2016rap,Cholis:2014lta}.}}
\item{If the Galactic Center Gamma-Ray Excess is generated pulsars, there is a probability of 0.7 that $\mathcal{O}(10^3)$ bright low-mass X-ray binaries would be observed in the Inner Galaxy (by INTEGRAL, etc.).}
\end{enumerate}

Given that no direct detection signals have appeared~\citep{LZ:2022lsv,XENON:2023cxc}, no pulsars have been found near the Galactic Center~\citep{Fermi-LAT:2023zzt}, and far too few low-mass X-ray binaries have been found in the Inner Galaxy~\citep{Haggard:2017lyq}, we can use this information to calculate new posterior probabilities, obtaining $P(T_{\rm DM}|D) \sim 0.65$ and $P(T_{\rm pulsars}|D) \sim 0.35$. In other words, taking this new information into account, I estimate that it is roughly 65\% likely that the Galactic Center Gamma-Ray Excess is generated by annihilating dark matter. This result is a reasonably accurate impression of my current thinking about this situation. I encourage you to state your own priors and use them to estimate the posterior probabilities that follow.

For the particle physicists reading this, I expect that many of you do not know much about pulsars, making it difficult for you to establish your related priors. Details aside, the most important information in this context is that, 1) pulsars are known to exist, and 2) they generate gamma-ray emission with a spectral shape that is similar to that of the Galactic Center Gamma-Ray Excess. The remaining points require some more specialized knowledge for which you might have to rely on the opinions of experts.\footnote{A pet peeve of mine is the phrase ``astrophysical sigmas,'' which is sometimes used by particle physicists to express the view that uncertainties quoted by astrophysicists are often unreliable. Although some astrophysical measurements suffer from large or difficult to quantify uncertainties, many do not. The same, of course, is true in particle physics. Consider, for example, the current state of interpretations of the muon's anomalous magnetic moment.} This is one reason why I think it is important for experts to clearly state their priors on a variety of propositions -- so that others can use that information to guide their own Bayesian reasoning.

From a dark matter perspective, this is an intriguing situation, but it is hardly conclusive. It certainly does not rise to the level of a discovery. To significantly increase the posterior probabilities in favor of either dark matter or pulsar interpretations of this signal, we will need one or more additional observations that are consistent with one of these hypotheses but not with the other. For example, if pulsars are responsible for this signal, they should also produce a flux of TeV-scale emission that traces the angular distribution of the Galactic Center Gamma-Ray Excess. By searching for this emission, the Cherenkov Telescope Array (CTA) will provide an important test of whether pulsars generate the observed excess~\citep{Keith:2022xbd}. In particular, given that $P(D_{\rm TeV}|T_{\rm pulsars}) \gg P(D_{\rm TeV}|T_{\rm DM})$, we can expect this observation to significantly aid in distinguishing between these hypotheses. 

Even more important to the question at hand will be future gamma-ray observations of Milky Way dwarf galaxies, which I consider to be the most likely way that this question will ultimately be resolved~\citep{Xu:2023zyz}. In the most recent analyses of Fermi data from the directions of dwarf galaxies, a small excess was identified. For dark matter with characteristics consistent with those required to generate the Galactic Center Excess, the statistical significance of this signal is approximately $\sim (1.5-2)\sigma$~\citep{McDaniel:2023bju}.\footnote{A similar analysis by \cite{DiMauro:2022hue} allowed for the possibility of spatially extended dark matter profiles around dwarf galaxies, and found a somewhat higher statistical significance for this signal.} To take this into account, I apply $P(D|T_{\rm pulsars}) \sim 0.13$ and $P(D|T_{\rm DM}) \sim 1$, obtaining $P(T_{\rm DM}|D) \approx 93\%$ and $P(T_{\rm pulsars}|D) \sim 7\%$. Despite being of only modest statistical significance, this information materially impacts the evaluation of my posterior probabilities. If this excess were to grow in statistical significance, even to $\sim$$3\sigma$ (corresponding to $P(D|T_{\rm pulsars}) \sim 0.003$), I would become quite confident in the conclusion that the Galactic Center Gamma-Ray Excess is generated by dark matter, $P(T_{\rm DM}|D) \approx 99.8\%$. Note that, in this context, a discovery does not necessarily require a posterior probability that is equivalent to the $p-$value corresponding to $5\sigma$, $1-(3\times 10^{-7})$.

\section{Closing Thoughts}

Since I have already indulged myself by preparing a very non-traditional contribution to these proceedings, I see no reason to not do so further. So, in that spirit, I will close by sharing with you some words from our universe's premiere physics-punk band:\footnote{\url{https://thespectraldistortions.bandcamp.com}}

\bigskip

{\it She holds her secrets closely

Her grip is strong like steel

And just when you think you've caught a glimpse

Of everything that's real, you ask

Can I trust myself?\\

She hides herself in darkness

Concealing all that's true

You think that you have seen the light

But many have been fooled before you, so you ask

Can I trust myself?}

\section*{Acknowledgements}
I would like to thank the organizers of the 182nd Nobel Symposium for inviting me to participate in that very interesting meeting. I would also like to thank the other participants of that symposium, whose discussions helped to inspire this contribution. This work has been supported by the Fermi Research Alliance, LLC under Contract No.~DE-AC02-07CH11359 with the U.S. Department of Energy, Office of Science, Office of High Energy Physics.

\bibliographystyle{elsarticle-harv} 
\bibliography{nobel}






\end{document}